\documentstyle[prl,multicol,aps]{revtex}

\input{epsf}
\tighten

\draft
\begin{document}
\title{ Bunching Transitions on Vicinal Surfaces and Quantum N-mers }
\author{V.~B.~Shenoy$^1$, Shiwei Zhang$^2$ and W.~F.~Saam$^3$\\
{\small $^1$Division of Engineering, Brown University, Providence, RI
02912.}\\
{\small $^2$ Department of Physics and Department of
Applied
Science, College of William and Mary, Williamsburg, VA 23187.}\\
{\small $^3$ Physics Department, The Ohio State University, Columbus,
OH 43210.}
}
\date{\today}
\maketitle

\begin{abstract}
We study vicinal crystal surfaces with
the terrace-step-kink model on a discrete lattice. Including
both a
short-ranged attractive interaction and a long-ranged repulsive
interaction
arising from elastic forces, we discover a series of phases in which
steps
coalesce into bunches of n steps each. The value of n varies with
temperature and
the ratio of short to long range interaction strengths. We propose that
the bunch
phases have been observed in very recent experiments on Si surfaces.
Within the
context of a mapping of the model to a system of bosons on a 1D lattice,
the bunch
phases appear as quantum n-mers.
\end{abstract}

\bigskip


\begin{multicols}{2}
\narrowtext

The study of the equilibrium properties of a stepped vicinal crystal
surface is
important from technological and fundamental perspectives. It also
provides a
fascinating example of a problem which has a much broader context,
reaching, via a
mapping onto a one-dimensional quantum chain, into the realm of 1D
quantum liquids.
In this Letter we address two related crystal surface problems.
The first is that of the
description of apparent
tricritical phenomena observed in a beautiful set of experiments by Song
{\it et
al}\cite{sm1} on
silicon surfaces, miscut away from the [113] direction towards [001].
The second is the related very recent observations of multiple-height
steps by Sudoh {\it et al}\cite{sudoh} on vicinal
silicon surfaces near the [113] crystalline
direction.

Recent theoretical work\cite{lassig,bhat} attempted to
explain the results of Song {\it et al} in terms of a continuum model of
steps
interacting via
a long-ranged repulsive elastic interaction and a short-ranged
attractive
interaction. This model fails to describe the observed bunching of steps
on the
vicinal surfaces coexisting with the (113) facet.
Here we explore the
consequences
of retaining the discrete, atomic nature of steps on a crystal surface
within the same model. We discover entirely different physics. The steps
do not phase separate but instead can coalesce into bunches whose size
depends on the relative strengths of the short- and long-range
interactions. Vicinal surface phases can be characterized by widely
separated n-bunches and transitions occur between phases having
different
values of n. We propose that bunch phases with large n correspond to the
two-phase coexistence region of Song {\it et al} and that the n=2,3,4
bunch
phases produce the multiple-height steps seen by Sudoh {\it et al}.

We employ the
terrace step kink model, which is a lattice model that can be mapped
onto an equivalent 1D quantum mechanical model of interacting spinless
fermions
or hard-core bosons\cite{jrs}.
 In the quantum picture it is
well-known that the
1-bunches form a Luttinger lattice liquid. Our results generalize this
picture to a
Luttinger lattice liquid which can form dimers (the 2-bunches), trimers
(the 3-bunches),
 and in general n-mers. The transitions between these phases
are quantum
many-body phenomena of a type hitherto unexplored.

Consider a crystal with lattice constants $(a_x,a_y,a_z)$, where the
steps are in
the $x-y$ plane and run, on the average, in the $y-$direction. In terms
of hard-core boson operators our Hamiltonian, for a system with $L$ lattice
sites in the $x-$direction, is written as
\begin{eqnarray}
{\cal H}& =&\eta(T)\sum_{i=1}^Ln_i
-t\sum_{i=1}^L\left[a_{i+1}^{\dagger}a_i +
a_{i+1}a_i^{\dagger}-2n_i\right]\nonumber\\
&&+ \sum_{j<i}\left[\frac{G}{|i-j|^2}
-U\delta_{|i-j|,1}\right]n_in_j,
\label{rsham}
\end{eqnarray}
where $t$, a monotonically increasing function of the temperature $T$,
is a hopping matrix element related to the step stiffness
$\Sigma$ by $\Sigma =\frac{(k_BT)^2}{2a_x^2t}$, 
$\eta(T)$ is the free energy per unit length of the steps, $G$
measures the strength of the elastic interaction\cite{noz}
between steps,
and $U$ gives a short range attraction between steps. Here $a_i$'s and
$a_i^{\dagger}$ are hard-core bosonic creation and annihilation
operators
respectively and
$n_i = a_i^{\dagger}a_i$ is the boson number operator on site $i$.
This representation is equivalent to the fermion representation
introduced in Ref.~\cite{jrs},
as demonstrated by application of the unitary Jordan-Wigner
transformation\cite{frad}. 
For $N$ bosons or steps confined on
$L$-sites
the ground state energy of $\cal{H}$, $E_0(s,T)$, is related to
the surface energy  of the vicinal  surface, $\gamma(s)$, through
$\gamma(s)/cos(\theta) = \gamma_0 + E_0(s,T)/La_x$, where $s=N/L$ is the
step or boson
density, $\theta = tan^{-1}(s)$ is the miscut angle
and $\gamma_0$ is the surface tension of the reference surface.

We now show that a number of features of this many body system can be
inferred from exact diagonalization of small systems in conjunction with
some simple analytical calculations. We find that for
sufficiently attractive interactions, the steps on the
surface rearrange themselves into bunches at low temperatures.
At $T=0$, where entropic effects are unimportant, the bunch size
is completely determined by the ratio of the strengths of the
attractive and repulsive interactions i.e. $U/G$.
 With increasing temperature,
the steps start peeling off from the bunches in a series of bunching
transitions, until only one step is left in a bunch
at high temperatures.
In what we call the extreme
 dilute limit, we obtain a phase diagram that shows these
bunching transitions as function of temperature.

Consider first the limit $t\rightarrow 0$, where the hopping part
of the Hamiltonian Eq.~(\ref{rsham}) can be completely
ignored; this is the
zero temperature limit  of the vicinal surface.
 In this limit, the energy of the system is obtained by  minimizing
the total potential energy,
and the minimum energy configuration consists
of bosons in bunches of size $n_b$, well separated from each
other.
In a bunch, the bosons sit next to one
another. For a bunch of size $n_b$, the energy per boson or free
energy per step is
given by
\begin{equation}
f(n_b,0)= \eta(0)
-\frac{U(n_b-1)}{n_b} +
\frac{G}{n_b}\sum_{i=1}^{n_b-1}\frac{(n_b-i)}{i^2},
\label{ener1}
\end{equation}
where we have neglected contributions from bunch-bunch interactions
that are smaller than the leading order terms given in Eq.~(\ref{ener1})
by a factor $s^2 \ll 1$. We
call the
limit in which these interactions are completely ignored, the {\em
``extreme
dilute limit''}(EDL). With increasing
$U/G$, the bunch size keeps increasing; for $U/G <1$ the bunch size is
$1$,
for $1 < U/G < 1.5$ the bunch size is 2, for $1.5 < U/G < 1.83$ the
bunch
size is $3$, for $1.83 < U/G < 2.08$ the bunch size is $4$, and so on.
 At the points $U/G = 1,1.5,1.83,2.08,...$
there
is a coexistence of bunches that differ in size by one. From plots
of
the energy per boson vs. $U/G$ for different $n_b$'s,
we find that the slopes of these curves at the points where they intersect are
not the same. The $T=0$ bunching transitions brought about by
changing $U/G$  can then be considered first order.

\begin{figure}
\epsfxsize=2.2in
\epsfysize=3.2in
\vspace{-0.8in}
\quad \leftline{\epsfbox{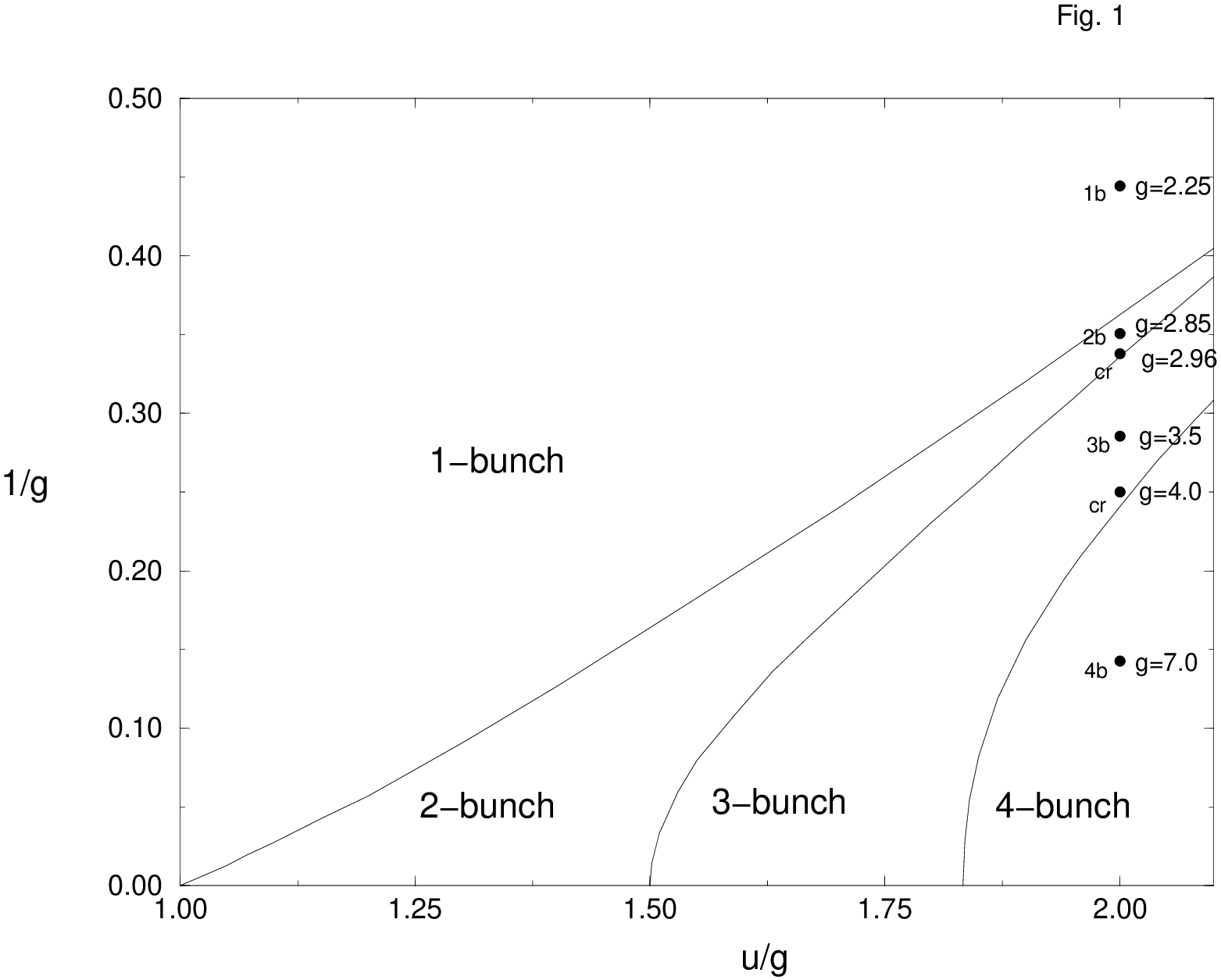}}
\vspace{0.1in}
\caption{ Phase diagram in the extreme dilute limit plotted in
$u/g$-$1/g$ space. Here $1/g$ plays the role
of temperature. For a given surface, the ratio of the strengths
of the short-range and long-range interactions fixes $u/g$, and
increasing the temperature corresponds to increasing $1/g$.
 The lines separate regions with stable bunch sizes
that differ by one as indicated in the figure.
 The points marked at $u/g=2$ correspond to the values of
$g$ for which the pair correlation functions, computed
using Green's function Monte Carlo (GFMC),
are shown in Fig.~\ref{pairc}. For each value
of $g$ we also indicate the bunch size (1b,2b,3b and
4b correspond to 1-,2-,3- and 4-bunches respectively,
while cr refers to the critical region) obtained from GFMC
simulations.}
\label{phasd}
\end{figure}

 At finite temperatures we expect the free energy per
step
in larger bunch sizes to
 decrease less rapidly compared to smaller bunches, as steps within
larger
bunches will have their fluctuations more constrained.
As a result,
larger bunches should rearrange themselves into smaller bunches,
with increasing temperature, until eventually at very
large temperatures only one step is left in the bunch. In order to
study
these bunching transitions, we introduce
$g=G/t$ and
$u=U/t$, which are the interaction strengths scaled by the hopping
matrix
element
$t$. Note that $g^{-1}$ plays the role of temperature. As in the $T=0$
case, we look at the free energy per step in the
$n_b$-bunch phase, $f(n_b,T)$, obtained by exactly diagonalizing
$\cal{H}$ for
$t\ne 0$. In
order
to obtain the phase boundaries between the $n_b$-bunch phase and
 the $n_b-1$ phase for a fixed $u/g$, one has to find the value of $g$
that satisfies the equation 
$f(n_b,T)-f(n_b-1,T)=0$. We do this numerically
by using a standard root finding algorithm which approaches
the root iteratively. The root finding algorithm
 requires repeated evaluation of $f(n_b,T)-f(n_b-1,T)$ using exact
diagonalization of the $n_b$- and $(n_b-1)$-bunches for different
values of $g$. The results are shown in Fig.~\ref{phasd}. A key point
emerging
from our calculations is
that, at the phase boundary between the
$n_b$- and $(n_b-1)$-bunch phases,
$\partial[f(n_b,T)-f(n_b-1,T)]/\partial T \ne 0$, implying 
that the
bunching
transitions are again first order in the EDL.

In performing the exact diagonilization calculations the 3 and
4-bunches were confined
 to $12$ sites on a ring, while for the 2-bunch we used 200
sites. This
is sufficient to get accurate answers because the bunches are tightly
bound with a
mean particle spacing between 1 and 1.5. 
We have also computed the
ground state energies using the Green's Function Monte Carlo (GFMC)
method\cite{GFMC}, which allows exact calculations for many-boson
systems, for bunches confined on lattices with up to 100 sites.
Our GFMC calculations, which we describe below, 
verified that the size effect in the 
exact diagonalization was negligible\cite{vs}.

We now consider the effects of including
bunch-bunch interactions on the phase diagrams obtained in the
EDL. Note that the contribution to the free energy per step
due to bunch interactions is smaller than the leading contribution
 $f(n_b,T)$ by a factor $s^2 \ll 1$ at all temperatures.
However,
in the region around the phase boundaries in Fig.~\ref{phasd}, called the
``critical region'',
the difference in the step free energy in the $n_b$ and $n_b-1$
phases is small. When the bunch-bunch interaction energy becomes
comparable to this difference, we have to explicitly include
it in the analysis in a non-perturbative way. Far from this
region the bunch interactions can be treated perturbatively.
When the bunch interactions are taken into account, we anticipate on
general
grounds that the degeneracy of the step free energies in the $n_b$ and
$n_b-1$
phases at the transition point is lifted, causing the
bunching transitions to become continuous.
The width of the critical
region, denoted by $\Delta
 T_{crit} \propto s^{p_{n_b}}$ ,
vanishes in the limit of vanishing densities (the EDL) as the bunch
interactions
approach zero in this limit.  
The exponent $p_{n_b}$, along with the shift in
bunching transition temperature from the EDL
due to bunch interactions, 
will be determined below.

 The bunch-bunch interaction energy in
regions away from the critical region can be computed
perturbatively
 by noting two key points: (1)The average spacing in a
bunch (for 2-,3- and 4-bunch phases in Fig.~\ref{phasd}, for $u/g<3$,
remains well under two lattice spacings\cite{vs},
so that the
steps in
a bunch remain tightly bound as a single
 entity. Since the bunches are very strongly
bound,
the stiffness of a bunch can then be approximated as
$\Sigma_b = n_b \Sigma$, where $\Sigma$ is
the
stiffness of a single step.
This expression for bunch stiffness is consistent with the experimental
observations of the stiffnesses of the double, triple and quadruple
steps
made by Sudoh {\it et al}\cite{sudoh}.
(2) The bunches interact with each other with a
 renormalized
inverse square potential of strength $g_b = n^3_bg$.  This
 approximation
is justified when the spacing
between the bunches is very large.
We note that the contributions to the energy of the system arising from
the short range bunch-bunch interactions are smaller
than the inverse square contribution by a factor proportional
to the density of steps $s \ll 1$. This can be shown
very easily using a Hartree-Fock estimation of the short-ranged
 contribution\cite{jrs}.
The energy of the system can then be computed using the
Calagero-Sutherland model
of hard-core bosons interacting via an inverse square law,
for which an exact solution was provided by Sutherland\cite{Suth}.
Using his
solution, we can write the energy per site in a system with
bunches of size $n_b$ as
\begin{eqnarray}
\frac{E_0[s,T,n_b]}{L} &= &{f(n_b,T)a_x\over
a_z}s\nonumber\\
&& + \left(\frac{{\pi}^2(k_BT)^2
\lambda_b^2 a_x}{6{n_b}^4a_z^3\Sigma }\right)s^3 + O(s^4),
\label{perte}
\end{eqnarray}
where
$
 \lambda_b =
\frac{1}{2}\left[1+\sqrt{1+2n_b^3g}\right].
$
Using this energy, one can compute such quantities as
surface stiffness and crystal shapes. These computations will appear
elsewhere.\cite{vs}

In the critical region, since the energies of both the $n_b$ and
$n_b-1$ bunches are nearly equal, we retain the terms of order
$s^3$ arising from bunch-bunch interactions. In order
to estimate the width of the critical region,
we use the criterion  $E_0(s,T,n_b-1) \approx E_0(s,T,n_b)$, to see that
$s^2 \sim |f(n_b,T)-f(n_b-1,T)|$ holds in the critical region.
 If $|f(n_b,T)-f(n_b-1,T)| \sim |T-T_c(n_b \rightarrow
n_b-1;0)|^{\alpha_{n_b}}$,
we obtain the result $p_{n_b} = 2/\alpha_{n_b}$, for $p_{n_b}$
introduced
earlier. For bunching transitions first order in the EDL, we have
$\alpha_{n_b}=1$ and therefore $p_{n_b}=2$.
As pointed out earlier, since all the bunching transitions shown in
Fig.~\ref{phasd} were found to be first order in the EDL, the width
of the critical region is $\Delta T_{crit} \propto s^2$. Extending
the phase diagram to larger values of $u/g$, we find that
for $u/g>2.4$\cite{vs},  the 1-bunch to 3-bunch  transition
preempts
the 1-bunch to 2-bunch transition and that for large $u/g$ all
transitions
asymptotically
approach the line $1/g=2(u/g)$. Further, all the bunching transitions
that the system undergoes for $u/g$ up to 8 were found to be first
order\cite{1-2t}.
Note next that the simplest conjecture for the shift of the
transition temperature due to bunch interactions is that it is of the same
order of magnitude as the width of the critical region.
Explicitly, for bunching transitions from the
$n_b$-bunch to the $n_b^{\prime}$-bunch phase that are first order in
the EDL,
we have
\begin{equation}
|T_c(n_b\rightarrow n_b^{\prime};s)-
T_c(n_b\rightarrow n_b^{\prime};0)| \propto s^2.
\label{Tshift}
\end{equation}

\begin{figure}
\epsfxsize=2.in
\epsfysize=2.5in
\vspace{-0.6in}
\quad \leftline{\epsfbox{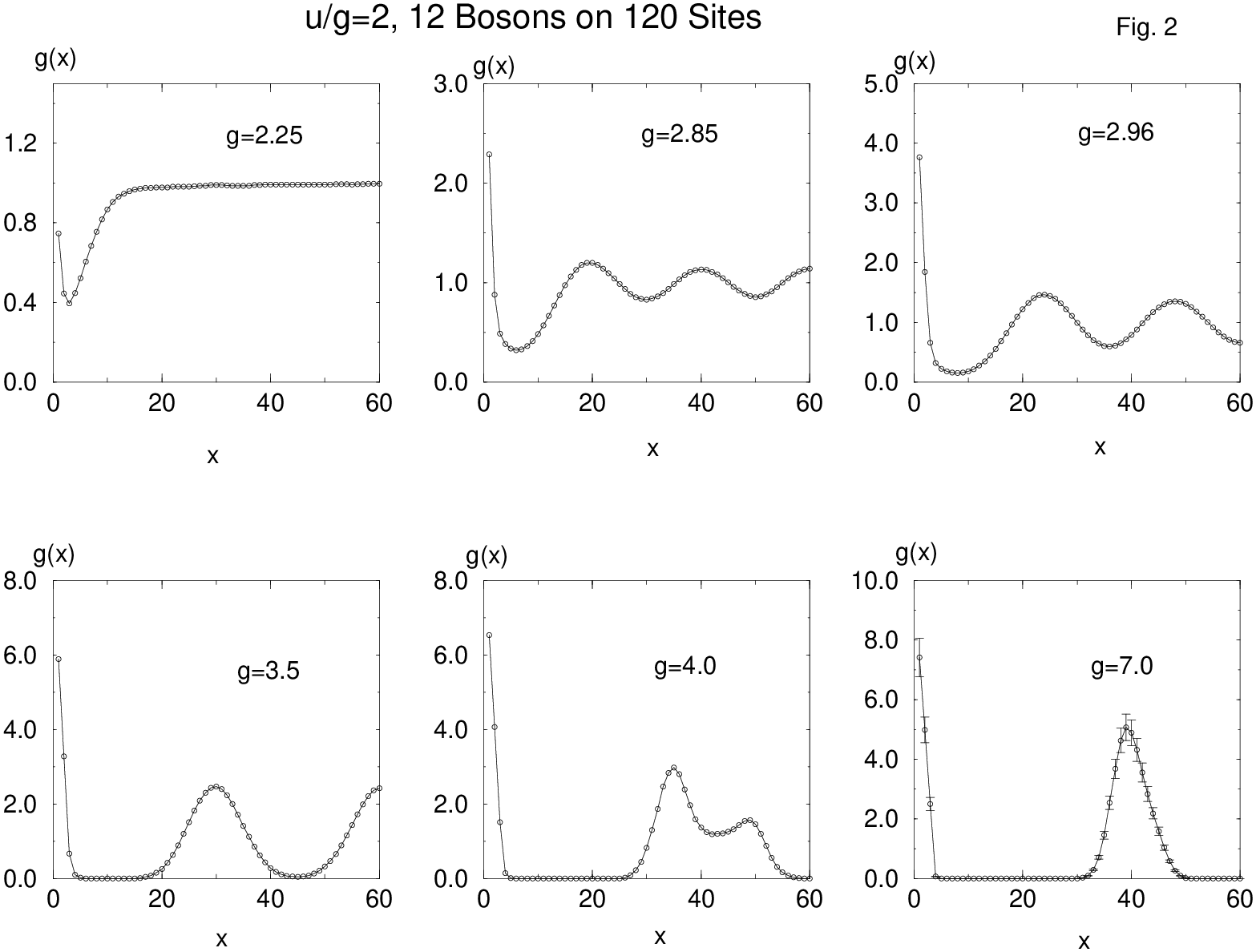}}
\vspace{0.2in}
\caption{Boson pair correlation function $g(x)$ (see Eq.~(\ref{corf})) for 12
bosons
on a ring with 120 sites. We have chosen  $u/g = 2.0$ and
plotted $g(x)$ for values of $g$ indicated in the
figure. The points are also marked in Fig.~\ref{phasd}  for comparison.
By counting the number of peaks in $g(x)$ and making use of the
periodic boundary condition the number of bosons in a bunch
can be established.
We see that there are 3 4-bunches for $g=7.0$, 4 3-bunches for $g=3.5$
and
6 2-bunches for $g=2.85$. For $g=2.25$  we have a completely unbound
1-bunch phase,  while
the points $g=4$ and $g=2.96$ are in critical regions. The 
statistical error
bars
have only been shown for the case $g=7$; they have 
similar magnitudes in all the other
cases.}
\label{pairc}
\end{figure}

We have also
performed GFMC simulations
 to study
the
physics of bunching transitions in the ``thermodynamic limit'' where
the bunch interactions are treated non-perturbatively and to verify
the results on bunching from exact diagonalization. 
To explicitly
illustrate the bunching transitions,
we compute the mixed estimator\cite{GFMC} of
 the pair correlation function, which for $N$
bosons on $L$ sites ($L$ even and $1\le x \le L/2-1$) defined by
\begin{equation}
g(x) = \frac{L}{2N^2}<\sum_{i \ne i^{\prime}} \delta\left(x-x_i +
x_{i^{\prime}}\right)>,
\label{corf}
\end{equation}
where  $x_i$ and $x_{i^{\prime}}$ are the coordinates
of the bosons $i$ and ${i^{\prime}}$ and the brackets indicate
the quantum average in the ground state. In Fig.~\ref{pairc} we
show plots of $g(x)$ for $u/g=2$. The figure vividly
demonstrates the transitions from the 1-bunch to 2-bunch, the 2-bunch to
3-bunch and the
3-bunch
to 4-bunch. For
comparison we have marked in the phase diagram in Fig.~\ref{phasd}
the bunch sizes for each
value of $g$ obtained by counting the number of peaks in $g(x)$.
The phases predicted for
12 bosons confined on 120 sites is in excellent agreement
with the EDL phase diagram obtained using exact diagonalization.
Note that for points very close to the phase boundaries ($g=4$ near
the 3-4 boundary and $g=2.96$ near the 2-3 boundary) we see
a continuous evolution from the $(n_b-1)$-bunch phase
to the $n_b$-bunch phase, consistent with lifting of level degeneracy 
as a result of the bunch-bunch interactions.
We have studied
the pair correlation functions for a number of other values of
$u/g$ to verify the accuracy of the phase diagram shown in
Fig.~\ref{phasd}\cite{vs}. Our GFMC energies 
agree with Eq.~(\ref{perte}).

Recently the effect of competing short and long range interactions
on
the finite temperature phase transitions on vicinal surfaces was
considered
by Lassig\cite{lassig} and Bhattacharjee\cite{bhat}, who used a
continuum
version of the model employed here. They analyzed the phase transition
using renormalization group (RG) techniques, finding that,
for a given $g$, when the short range attraction is
sufficiently attractive, the steps
tend to phase separate. This
picture is
fundamentally different from ours. In our picture, the steps do
not phase
separate. Instead, they form separated bunches of a finite size. The
bunches then
interact with each other with renormalized interaction
parameters as
discussed above. With increasing temperature
the steps undergo a series of peeling transitions where the
bunch size progressively decreases.
The size of a bunch and peeling transitions are crucial
effects of a physical region where the continuum model is not valid.

In the experiments of Song {\it et al}\cite{sm1},
above a temperature $T_c(s)$, the surface is
uniformly stepped, while below
$T_c(s)$ it consists of facets separated by
bunched steps.
The number of steps in a bunch was seen to be about $22$. From the
measurements it was inferred that $s \sim |T_c(0)-T_c(s)|^{\beta}$,
with $\beta = 0.42 \pm .1$.
Our picture suggests that we identify the experimentally observed
transition
as a 1-bunch to 22-bunch transition.
With our simple choice of potential, this particular transition does not
appear.
However, we do observe, for $u/g>2.4$\cite{vs},
that the 1-3
transition preempts
the 1-2 transition. For a potential with next-nearest neighbor
attractive interactions of sufficient strength, 
even at $T=0$,
a 1-bunch directly goes to a 3-bunch, bypassing the
2-bunch. For potentials with more complicated features, it can be
expected that a 1-22 transition will occur.
For quantitative 
progress here, a more detailed understanding
of
the step-step interaction on Si(113) surfaces is required.

 From Eq.~(\ref{Tshift}), if a 1-22 transition 
is first order in the EDL, {i.e.}, the
free energy
curves $f(22,T)$ and
$f(1,T)$ intersect with a finite slope, 
the shift in transition temperature scales like $s^2$. It then
follows that $\beta = 0.5$, in agreement with the
experimentally observed exponent. It is important to note that in our
picture $\beta$
describes a curve of continuous 1-22 bunch transitions rather than a
curve of
first order transitions associated with a tricritical point. Within the
recent RG
calculations,
$\beta$ is non-universal, requiring a
particular choice of $u/g$ to produce agreement with experiment. In
contrast, our
result $\beta=0.5$ is robust, applying to any transition that is first-order in the
EDL. All
transitions in our model are indeed first-order in the EDL. It is precisely
in this
situation that the RG calculations fail.

It remains difficult to
reconcile some observed features of the bunch period and misorientation
angle  with our
picture. In particular, the minimum observed in the bunch period as a
function of the misorientation
angle is readily understood\cite{sm1} within a picture, due to
Marchenko\cite{mar2}, where
coexistence of two macroscopic facets is unstable and replaced by a
periodic groove structure which
is in fact a bunch phase. The difference between the Marchenko picture
and ours is that the bunch size
in the former is determined by a balance between edge energies and
strain energies, whereas in the
latter the bunch size is determined by the competition between short-
and long-range step
interactions. Further research, extending our work to the case
of large
bunches not necessarily widely separated, is required to elucidate the
connection
between these two pictures.

Our interpretation of the results of Song {\it et al} gains
considerable support
by the direct observation of single, double, triple, and quadruple steps
by Sudoh
{\it et al}. The bunch sizes in the two experiments are expected to
differ
because the vicinal surfaces involved have different orientations with
respect
to the [113] direction, leading to different short and long range
step-step
interactions. Further exploration of the vicinal surfaces of Si near the
[113]
direction in an effort to fully elucidate the connections between the
two sets of
experiments will be crucial to the further development of this area.

\end{multicols}


\begin{references}

\bibitem{sm1} S. Song and S. G. J. Mochrie, {\it Phys. Rev.
Lett.} {\bf 73}, 995 (1994); S. Song and S. G. J. Mochrie,
{\it Phys. Rev. B} {\bf 51}, 10068 (1995); S. Song {\it et al},
{\it Surf. Sci.} {\bf 372}, 37 (1997).

\bibitem{sudoh} K. Sudoh {\it et al}, {\it Phys. Rev. Lett.} {\bf 80},
5152 (1998).

\bibitem{lassig} M. Lassig, {\it Phys. Rev. Lett.} {\bf 77}, 526 (1996).

\bibitem{bhat} S. M. Bhattacharjee, {\it Phys. Rev. Lett.} {\bf 76},
4568 (1996).

\bibitem{jrs} C. Jayaprakash, C. Rottman, and W. F. Saam, {\it Phys.
Rev.
B} {\bf 30}, 6549 (1984).

\bibitem{noz} V. I. Marchenko and A. Ya. Parshin, {\it Sov. Phys. JETP},
{\bf
52},
129 (1981).

\bibitem{frad} E. Fradkin, {\em Field Theories Of Condensed Matter
Systems},
(Addison-Wesley, Redwood City, California, 1991).

\bibitem{GFMC}
See, e.g., D.~M.~Ceperley and M.~H.~Kalos, in {\it Monte Carlo Methods in
Statistical Physics\/}, ed.\ by K.~Binder (Springer-Verlag,
Heidelberg, 1979); N. Trivedi and D. M. Ceperley, {\it Phys. Rev. B}
{\bf
41},
4552 (1990).


\bibitem{vs} V.~B.~Shenoy, Shiwei Zhang and W.~F.~Saam, to be
published.

\bibitem{Suth} B. Sutherland, {\it J. Math. Phys.} {\bf 12}, 246, 251
(1971).


\bibitem{1-2t} We find that, in the EDL, the 1-2 transition is first order for
$g>1.5$, while
it becomes continuous for $g<1.5$. However the system does not
undergo this continuous transition since it is preempted by the 1-3
transition for $g<1.86$ and $u/g > 2.4$ as noted in the text.

\bibitem{mar2} V. I. Marchenko, {\it Sov. Phys. JETP} {\bf 54}, 605
(1981).

\end{references}
\end{document}